\documentclass[12pt]{iopart}

\usepackage{graphicx}
\begin{document}

\article[Ionic-liquid-like local structure in a $LiNO_3-Ca(NO_3)_2-H_2O$]{}
{Ionic-liquid-like local structure in $LiNO_3-Ca(NO_3)_2-H_2O$ as studied by ion and solvent nuclei NMR relaxation}
\author{Vladimir V Matveev$^{1,3}$, Petri Ingman$^2$, Erkki L$\ddot{a}$hderanta$^3$}\address{$^1$ Department of Quantum
Magnetic Phenomena, Physical Faculty, St.-Petersburg State University, RU-198504 St.-Petersburg,
Russia}\ead{\mailto{vmatveev@nmr.phys.spbu.ru}}
\address{$^2$ NMR Laboratory, Department of Chemistry, University of Turku,
FI-20014 Turku, Finland}
\address{$^3$ Laboratory of Physics, Lappeenranta University of Technology,
Box 20, 53851 Lappeenranta, Finland}
\begin{abstract}
Relaxation rates in the $13mLiNO_3-6,5mCa(NO_3)_2-H_2O$ ternary system have been measured for nuclei of water ($^1H$ and
$^{17}O$), anion ($^{14}N$), and both cations ($^7Li$, $^{43}Ca$). The data analysis reveals the system structure as
consisting of two main charged units: [Li(H$_2$O)$_4$]$^+$ and [Ca(NO$_3$)$_4$]$^{2-}$. Thus the system presents
inorganic ionic liquid like structure.
\end{abstract}
 \maketitle
\normalsize
\section{Introduction}
Ionic Liquids (ILs), i.e. room temperature molten salts attract growing attention as
 very promising systems both for academical knowledge and for industrial applications \cite{IL1,IL2,IL3,IL4,IL5}.
 Therefore, investigation of their structure and dynamics by various methods is nowadays one of the leading topic in
 Physics and Chemistry of liquid systems. There are a number of NMR papers on ILs at the moment (see e.g.
 \cite{rev1,g1,g2,g3,g4,g5}). However, they are still far from using the full potential of multinuclear NMR technique.
 During few last decades  the technique has shown its high efficiency for various electrolyte solutions, see
 \cite{Holz-1986,Kowal,Chizhik-MP,my-rev} and references within.

 Concentrated electrolyte solutions could be good models for ionic liquids due to clear similarity of the systems, and in
 this connection other potentialities of NMR might be demonstrated using these easer prepared samples.
 Here we report NMR study for the
$13mLiNO_3\--6.5mCa(NO_3)_2\--H_2O$ ternary system where multinuclear relaxation data allow evaluation of the mobility of
all nuclei both in solvent and solute.

NMR technique is an important information source on molecular mobility in neat ILs as well as in various electrolyte
solutions as the NMR relaxation rates under certain experimental conditions are proportional to characteristic times of
translational and/or rotational motion of ions and solvent. In the investigated
system there are two different channels of spin-lattice relaxation which control the spin-lattice relaxation rate,
$T{_1}^{-1}$ (later denoted as R for simplicity). For $^{14}N$, $^{17}O$, and $^{43}Ca$ nuclei the main contribution is
due to quadrupolar interaction, and corresponding $T_{1Q}^{-1}$ is given by \cite{Abraham,Kowal}
\begin{equation}
\frac{1}{T_{\rm 1Q}}= \frac{3\pi^2}{10}\frac{2I+3}{I^2(2I-1)}\nu{_{\rm Q}}{^2}\tau_{eff}.
\end{equation}
Here $\tau_{eff}$ is the effective correlation time,  in many cases equal to the rotation reorientation time, $\tau_{r}$,
of ion/molecule which contains the investigated nucleus; $I$ is the spin of the nucleus; the quadrupolar coupling
constant, $\nu_Q=e^2qQ/h$ where $eq$ is the electric field gradient at the nucleus and $eQ$ is the nucleus quadrupole
moment. Equation (1) is valid only for nuclei with axial symmetry of the quadrupolar tensor.
Another mechanism of relaxation, mostly relevant here for $^1$H nuclei, is magnetic or dipole-dipolar spin-spin
interaction. Corresponding relaxation rate, $T_{\rm 1m}^{-1}$ is proportional to the correlation time, $\tau_{c}$
\cite{SB-1948,Abraham}:
\begin{equation}
\frac{1}{T_{\rm 1m}}= A \tau_{c}
\end{equation}
where $\tau_c$ is, similar to $\tau_{eff}$, typically equal to $\tau_{r}$; $A$ can be generally expressed in the form: $$
A = \gamma_I^2\gamma_S^2\hbar^2r^{-6}$$ for two different nuclei $I$ and $S$ and $$ A = (3/2)\gamma_I^4\hbar^2r^{-6}$$
for two equal nuclei $I$. Here $\gamma_I$ and $\gamma_S$ are the nuclear gyromagnetic ratios for nuclei $I$ and $S$
correspondingly, $r$ is the internuclear distance. Both (1) and (2) are valid under the simplifying assumption of
so-called "extreme narrowing" condition \cite{Abraham,Kowal}
\begin{equation}
\omega_0\tau_c\ll1.
\end{equation}

For $^7$Li coexist dipole-dipolar and quadrupolar relaxation channels and their relative contributions depend on
molecular environment of the ion.

As it is evident from (1) and (2)  the relaxation rate is proportional to molecular correlation time and allows an
evaluation of the $\tau_{r}$ value from NMR relaxation data. However, an absence of precise magnitudes of $\nu_Q$  and
$r$ prevent this evaluation in a number of cases. Especially it it true for concentrated solutions with unknown local
structure of the ions and solvent environment. To overcome or at least to reduce this difficulty it is useful to explore
molecules with fixed chemical environment (chemical structure) and with internal $\nu_Q$ value for quadrupolar nucleus.
In such kind of samples one may assume an approximate conservation of $\nu_Q$ or $r$ value and hence, an approximate
constancy of the coefficients (factors) in (1) and/or (2). Then an increase of the relaxation rate reveals an increase of
$\tau_{r}$ and allows a semi-quantitative comparison of solutions with different electrolyte concentrations at least in the
form of the  ratio of $\tau_{r}s$. For the investigated system such nuclei are $^{14}$N in NO${_3}{^-}$ anion where
$\nu_Q$ value is affected primarily by the nitrogen -- oxygen chemical bond and $^1$H where $r$ is controlled by
hydrogen--hydrogen distance in water molecule. For both these cases it is reasonable to assume that the coefficients in (1)
and (2) are virtually independent of salt concentration. This assumption is used below to describe the main trends in
relaxation rates. 
\section{Experimental procedures}
The ternary system under study was prepared as described earlier \cite{MK}. Spin-lattice relaxation times, $T_1$, for
$^1$H and $^7$Li nuclei were measured using a common inversion-recovery ($180\--\tau\--90$) technique at 8~MHz resonance
frequency  with a home-made relaxometer. The spectral linewidth, $w$, for $^{14}$N, $^{17}$O and  $^{43}$Ca nuclei were
measured with Bruker AM-500 and AVANCE-400 spectrometers at magnetic field of 11.8~T and 9.4~T, correspondingly.
Spin-spin relaxation time, $T_2$, was then calculated according to the common $T{_2}$~=~($\pi w){^{-1}}$ relation. This
was used instead of $T_1$ as the "extreme narrowing" condition (3) was valid for these nuclei. All measurements were
carried out at room temperature.
\section{Results and Discussion}
Experimental relaxation rates for all nuclei in the mixture are collected on the
Table~1.  Also are included corresponding literature  values, $R_0$, in pure water or in proper dilute solution with
calculated $R/R_0$ ratios. As it is evident from (1) the $R/R_0$ values are not unambiguously proportional to increasing
of $\tau_{r}$. But even a qualitative analysis of the Table~1 allows one to assume more or less clear picture of the
short range order and dynamics in our system. First, the data reveal close $R/R_0$ values for $^1$H and $^{17}$O water
nuclei. In the case of $^{17}$O it is reasonable to suggest, that $\nu_Q$ depends mainly on the chemical bond strength in
the water molecule and remains more or less constant in the solution. Hence, an increasing of the relaxation rates for
these nuclei reflects primarily an increase of $\tau_{r}$ due to the solute concentration increasing. The same estimation
looks valid for proton relaxation due to its mainly intramolecular magnetic origin. Thus, the measurements show that the
rotational correlation time for water molecule in the system exceeds that in pure water approximately by a factor of 20.

For $^7$Li nucleus the $R/R_0$ value is close to $^1$H and $^{17}$O ones for water. It is well known that the $^7$Li
relaxation rate is the sum of the dipolar (magnetic) and quadrupolar terms \cite{Abraham,rev1}. In the dilute aqueous
solution it has been shown that these two contributions are practically equal \cite{Mazitov}. However, some additional
assumption is required to separate them in the studied solution as the coefficients in (1) and (2) could vary with
concentration. Nevertheless, close $R/R_0$ values for $^7$Li cation and for water allow one to assume a conservation of
$^7$Li dipolar to quadrupolar contribution ratio for this system and to attribute the R/R$_0$ increase to $\tau_{r}$
increasing. In any case an increasing of $\tau_{r}$ for $^7$Li relaxation is much less in comparison with other ions
$R/R_0$s. In fact, for $^{14}$N and $^{17}$O nuclei in the $NO{_3}{^{-}}$ anion $R/R_0$ values exceed those for dilute
solutions by a factor of $n\cdot10^2$, and the same is observed for $^{43}$Ca nucleus of Ca$^{2+}$ cation.

For $^{14}$N and $^{17}$O nuclei in the anion their $\nu_Q$ values are affected primarily by the nitrogen -- oxygen
chemical bond. However, equation (1) is valid only for $^{14}$N nucleus whereas $\tau_{eff}$ for $^{17}$O is not equal to
$\tau_{r}$ for the anion; for more detail see \cite{NO3-1,NO3-2} and references within. Nevertheless, $R/R_0$ for
$^{14}$N reveals unambiguously a huge increase of the anion $\tau_{r}$. The environment of Ca$^{2+}$ cation  has a high
symmetry in dilute aqueous solutions and therefore, an increase of its relaxation rate is a product of two effects,
namely (i) an increase of the correlation time for Ca$^{2+}$ cation and (ii) a distortion of symmetry of the cation
leading to increase of $\nu_Q$. As a result the $^{43}$Ca relaxation rate displays the maximal $R/R_0$ value.
\begin{table}
\caption{\label{arttype}Relaxation rates of the nuclei in the investigated
solution.} \footnotesize\rm
\begin{tabular*}{\textwidth}{@{}l*{15}{@{\extracolsep{0pt plus12pt}}l}}
\br
nucleus;  ion/solvent&R, s$^{-1}$&R$_0$, s$^{-1}$&R/R$_0$\\
\mr
$^1$H;  H$_2$O&4.95&0.29\cite{SB-1948,Chizhik-MP}&17.1\\ $^7$Li;
Li$^+$&1.04&0.055$\pm$0.005\cite{Li-1,Mazitov}&19$\pm$2\\ $^{17}$O;  H$_2$O&2.65
10$^3$&138$\pm$4\cite{17O-1,17O-2}&19.2\\ $^{17}$O;  NO${_3}{^-}$&32,7 10$^3$&317\cite{NO3-1}&103\\ $^{14}$N;
NO${_3}{^-}$&1.58 10$^3$&7.69\cite{NO3-1}&205\\ $^{43}$Ca;  Ca$^{2+}$&198$\pm$18&0.80\cite{Ca-0}&240$\pm$20\\
\br
\end{tabular*}
\end{table}

\textbf{The analysis above may be summarized as follows:}
\begin{enumerate}
\item An increase of relaxation rates for $^{14}$N nucleus in the anion and
$^{43}$Ca nucleus in calcium cation is much more pronounced than that for $^7$Li nucleus in lithium cation and for
both $^{1}$H and $^{17}$O nuclei in water molecule.
\item The anion $\tau_{r}$ value obtained from the $^{14}$N $R/R_0$ under the
assumption of $\nu_Q \approx const$ is well over $\tau_{r}$ value for Li$^+$ cation which can be obtained from $^7$Li
$R/R_0$ with any reasonable $\nu_Q$ value.
\item water and Li$^+$ $\tau_{r}$  values are nearly the same.
\end{enumerate}
In other words this analysis leads to the conclusion that Li$^+$ cation is surrounded by water only, whereas Ca$^{2+}$ cation and NO${_3}{^-}$ anions combine themselves into some complex anions. The stoichiometry of the solution, Ca: 2Li: 4NO$_3$: 8.5H$_2$O, allows the almost complete distribution of solvent and anions among the cation surroundings. Therefore it is reasonable to assume that the local structure of the system consists of two main structural units or clusters:
$$[Li(H_2O)_4]^+$$ and $$[Ca(NO_3)_4]^{2-}.$$ The units are in 2:1 proportion and coulomb interaction determines mainly
the middle\--range order and dynamics of the system. It means that the electrolyte solution under investigation presents a
kind of inorganic ionic liquid, and it is the first example of such kind, at least as far as the authors' knowledge.
\section{Conclusion}
We have demonstrated some possibilities of multinuclear NMR technique to obtain information on local structure and
dynamics of solvent and ions in a complex system consisting of water molecules and a number of counterions  --
concentrated electrolyte solution. The data obtained allow a description of the local structure of the system as a
composition of two charged units \-- complex counterions \-- and lead to estimation of dynamics of solvent and ions. The
same methodology can be used for investigation of properties of various ILs in neat state as well as in a mixture with
solvent.
\ack We gratefully acknowledge Prof Maria K. Khripoun for the sample. We also  acknowledge Prof Sergei V. Dvinskikh and
Prof Viktor I. Tarkhanov for fruitful discussions. V.V.M. much indebted to Ivan S. Podkorytov for his help during working
at AM-500 spectrometer. The work was partly supported by Russian RFBR grant \#07-08-00548.

\end{document}